%% file: main.tex
\documentclass[conference,letter]{IEEEtran}
\IEEEoverridecommandlockouts

\usepackage{color}
\usepackage{amsthm}
\usepackage{mathtools}

\DeclarePairedDelimiter\floor{\lfloor}{\rfloor}
\usepackage{bbm}
\usepackage{amssymb}
\usepackage{tikz}
\usepackage{comment}
\usepackage{enumitem}
\usepackage[ruled,vlined, linesnumbered]{algorithm2e}

\SetCommentSty{mycommfont}

\RestyleAlgo{ruled}
\SetInd{0em}{0.5em}
\SetKwComment{Comment}{/\!\!/}{}
\usepackage{booktabs}
\usepackage{cite}
\usepackage{steinmetz}
\usepackage{subcaption}
\usepackage{dblfloatfix}

\newcommand\blfootnote[1]{%
  \begingroup
  \renewcommand\thefootnote{}\footnote{#1}%
  \addtocounter{footnote}{-1}%
  \endgroup
}

\usepackage{caption}
\input{others/com.tex}

\begin{document}
\bstctlcite{IEEEexample:BSTcontrol}

\title{Hybrid Beam Alignment for Multi-Path Channels: A Group Testing Viewpoint}
\author{Ozlem Yildiz, Abbas Khalili, Elza Erkip\\
NYU Tandon School of Engineering,\\
Emails: \{zy2043, ako274, elza\}@nyu.edu}

\maketitle

\begin{abstract}

High-frequency bands such as millimeter-wave and terahertz require narrow beams due to path loss and shadowing. Beam alignment (BA) methods allow the transceivers to adjust the directions of these beams efficiently by exploiting the channel sparsity at high frequencies. This paper investigates BA for an uplink scenario, where the channel between the user equipment (UE) and base station (BS) consists of multiple paths.
The BS wishes to localize the angle of arrival of each of these paths with a given resolution using the least number of time slots. At each time slot of the BA, the UE transmits a BA packet and the BS uses hybrid beamforming to scan its angular region. To minimize the expected BA duration, a group testing framework is devised, and the associated novel analog and hybrid BA strategies are described. Simulation studies show the performance improvement both in noiseless and realistic 5G mmWave BA settings.

\blfootnote{This work is supported by
National Science Foundation grants EARS-1547332, and
SpecEES-1824434}
\end{abstract}

\vspace*{-0.4cm}
\section{Introduction}

To compensate for the ever-growing demand for data rate, next-generation wireless networks are envisioned to operate at high frequencies such as millimeter wave (mmWave) and terahertz (THz), enabling multi-Gbps throughputs through the use of the available spectrum at these frequencies  \cite{mmWave-survey-nyu} \cite{xing2021propagation}. Communication in high frequencies faces obstacles such as high path loss and shadowing which can drastically degrade the performance \cite{mmWave-survey-nyu}, necessitating the use of directional beams (a.k.a. \textit{narrow beams}) for communication \cite{kutty2016beamforming}.


Experimental results in mmWave \cite{akdeniz2014millimeter} and THz frequencies \cite{xing2021propagation} demonstrate that the communication channels are sparse and consist of a few (up to four) spatial clusters. Jain et al.  \cite{jain2021two} show that locating multiple clusters improves reliability and throughput. Multiple beams can also be used for recovery under blockage which is often observed at higher frequencies \cite{ganji2021overcoming}. Devising efficient beam alignment (BA) techniques to identify the channel clusters is important for mmWave and THz communications. In BA, a wireless transceiver searches its angular space using a set of scanning beams to localize the direction of the channel clusters, namely, the angle of arrival (AoA) and angle of departure (AoD) of the channel clusters at the receiver and transmitter sides, respectively.

BA methods can be categorized into different classes, such as \textit{interactive} or \textit{non-interactive}, and \textit{hybrid} or \textit{analog} BA. To elaborate, let us consider the case of one UE and one BS, where the UE transmits a BA packet every time slot and the BS tries to localize the AoAs of the channel. In non-interactive BA, at each time slot, the BS uses a set of predetermined scanning beams that are independent of the measurements in prior time slots. In interactive BA, the BS uses the prior scanning beams' measurement results to refine future scanning beams and better localize the AoA of the channel. In analog BA, the BS uses only one radio frequency (RF)-chain corresponding to one scanning beam at each time slot. In hybrid BA, however, the BS uses multiple RF-chains resulting in the simultaneous use of multiple scanning beams. 

%
Khalili et al. \cite{khalili2021single} consider interactive analog BA for a single-user, single-path scenario, where the resulting feedback is delayed.
Chiu et al. \cite{chiu2019active} study noisy interactive analog BA for single-path channels to find a data beam with a target resolution. Aykin et al. \cite{aykin2019multi} propose multi-lobe beam search (MLBS) for noiseless analog interactive BA and investigate its performance in the presence of noise. Noh et al. \cite{noh2017multi} develop a BA method for noisy hybrid interactive BA when the channel is single-path and Song et al. \cite{song2019efficient} provide a method for noisy hybrid non-interactive BA when the channel has multiple paths. 
The connection between group testing (GT) and analog non-interactive BA for multiple paths was investigated in \cite{suresh2019}. Unlike the analog BA for finding one strong path which is well investigated in the literature, less is known about the hybrid BA for locating multiple paths between a UE and a BS.

In this paper, we consider the problem of \textit{hybrid interactive BA} in an uplink scenario, where the channel between the UE and BS consists of \textit{multiple spatial clusters} (paths), and the BS is equipped with $N_{\rm RF}$ RF-chains. At each time slot of the BA, the UE transmits a BA packet with an omnidirectional transmit pattern while the BS scans its angular space using $N_{\rm RF}$ beams. 
The objective is to locate the AoAs of the channel clusters within the target beamwidth $\omega$ in the shortest BA duration.


Our strategies are based on GT, where the general goal is to identify a small number of defective items from a large group by the use of pooled testing \cite{hwang1972method}.



A summary of our contributions are as follows:

\begin{itemize}
    \item 
    We provide a duality between BA for identifying multiple paths and GT and devise an analog interactive BA algorithm accordingly (Sec.~\ref{subsec:analogba}). 
    We compare the expected BA duration of the proposed algorithm with the state-of-the-art multi-path analog BA methods and demonstrate the performance improvement (Sec.~\ref{sec:sim}).

    \item  We extend the proposed GT-based analog BA method to the hybrid scenario with multiple RF-chains by providing three different generalizations (Sec.~\ref{subsec:hybridba}). We show that the expected BA duration for a narrow target beamwidth can be reduced by a factor of two compared with our analog GT-based BA, and by more than a factor of five compared with a hybrid version of the state-of-the-art exhaustive search (Sec.~\ref{sec:sim}).

    \item We provide 5G mmWave simulations to assess the effectiveness of the proposed BA techniques in a realistic environment. In practice, the BS needs to determine the received SNR threshold to categorize the scanning results. We simulate the expected number of detected paths and outage probability of the algorithms for different threshold values to determine the optimum threshold. Then,
    we investigate the expected BA duration for the optimized threshold.
    Compared with the state-of-the-art, we observe that GT-based interactive hybrid BA improves the expected BA duration while having comparable expected number of paths and outage probability (Sec.~\ref{sec:sim}).

\end{itemize}

\section{System Model and Preliminaries} 
\label{sec:sys}
\subsection{Network Model}
\label{subsec:net}
We consider an uplink communication scenario, where the BS performs BA. We assume that the channel between the UE and the BS consists of $M$ paths, typically upper bounded by four with an average of two in mmWave and THz bands \cite{mmWave-survey-nyu, xing2021propagation}. We assume that while the BS scans its angular space using a set of scanning beams to determine the AoAs of the UE channel,  the UE transmits a BA packet every time slot  with an omnidirectional transmission pattern.

 Similar to \cite{aykin2019multi}, we denote the AoAs corresponding to the M paths by $\psi_m,~m\in{[M]}$ which are i.i.d and uniform between $[0, 2\pi)$ and unknown to the BS. We consider that the BS performs hybrid BA during which it simultaneously uses $N_{\rm RF}$ scanning beams at each time slot, where $N_{\rm RF}$ is the number of RF-chains at the BS. Given a fixed beamwidth, $\omega$,  the BS's goal is to find the angular intervals with width $\omega$ that include AoAs using the fewest number of BA time slots \cite{aykin2019multi, suresh2019}.


\subsection{Beam Alignment and Problem Formulation}
\label{subsec:BA}

We consider an interactive BA scenario in which the BS determines the next $N_{\rm RF}$ scanning beams at each time slot based on the measurements corresponding to previous scanning beams. We assume that the BS can determine if a past scanning beam includes a path or not before the next BA time slot \cite{aykin2019multi, suresh2019}. If a scanning beam includes at least one path, or equivalently, if the scan results in energy detection, we denote this by an acknowledgment (ACK), otherwise, it is denoted by a negative ACK (NACK).

We denote the hybrid interactive BA procedure by ${\textit{S}} (\omega, M, N_{\rm RF})$, which describes the $N_{\rm RF}$ scanning beams to be used at each time slot based on the previous scanning beams and their received ACK/NACK sequences. Here, $\omega$ denotes the angular width of the resulting data beam at the end of BA which includes $M$ AoAs. We denote the resulting expected BA duration as $E[T_{{\rm BA}}(\mathbf{\Psi, S})]$,where $\mathbf{\Psi} = (\psi_1, \psi_2, \ldots, \psi_M)$  is an AoA realization, $T_{{\rm BA},S}(\mathbf{\Psi})$ is the BA duration of $\textit{S}$  for given  $\mathbf{\Psi}$ and the expectation is taken over $\mathbf{\Psi}$. 

Our goal is to devise an interactive hybrid BA procedure ${\textit{S}^*} (\omega, M, N_{\rm RF})$ that minimizes the expected BA duration as in
\begin{equation}
    S^* ( \omega , M, N_{\rm RF}) = \argmin_{S( \omega , M, N_{\rm RF})} E[T_{{\rm BA},S}(\mathbf{\Psi})].
\end{equation}
Similar to \cite{hussain2017throughput,khalili2020optimal, aykin2019multi}, we first develop BA procedures assuming there is no noise in the BS detection of the paths (Sec.~\ref{sec:PF}) and investigate the impact of more realistic channel model including noise through 5G simulations (Sec.~\ref{sec:sim}).

\vspace*{-0.2cm}
\section{Proposed Methods}
\label{sec:PF}

We first provide a connection between the GT and BA problems using which we develop GT-based interactive BA methods for the hybrid BA optimization problem (in Sec.~\ref{sec:sys}).

\subsection{Group Testing and Beam Alignment}

GT is a well-known method to find $M$  defectives among a large number of items, $N$, using a small number of tests \cite{hwang1972method}. This is done by repeated testing where each test pools a subset of items and checks whether there is a defective item in the subset. A common assumption in the GT literature is that each test gives a binary answer: yes (at least one defective item among the ones tested) or no (no defective items). A GT method is classified as \textit{interactive} if each test is designed based on prior tests and their results, otherwise, \textit{non-interactive} \cite{aldridge2019group}.

In this paper, we propose to view the BA problem as a GT problem. To elaborate, consider the setup discussed in Sec.~\ref{sec:sys}. Given the desired final beamwidth $\omega$, we can divide $[0,2\pi)$ into $N= \frac{2\pi}{\omega}$ equal width angular intervals. When the channel has $M$ AoAs, up to $M$ of these $N$ angular intervals contain AoAs. As a result, the BA problem can be formulated as a GT problem in which we want to find up to $M$ defectives (angular intervals that include AoAs) among $N$ items (number of angular intervals) using a set of binary (ACK/NACK) tests (scanning beams). This analogy allows us to incorporate GT ideas and strategies and modify them as appropriate in the context of BA, particularly when multiple RF chains are used. Using this perspective, we first provide a GT-based interactive BA strategy when $N_{\rm RF} = 1$  (a.k.a analog BA) and then extend the algorithm to the case of $N_{\rm RF} = 2$ and provide multiple methods to generalize analog BA to hybrid BA. Our designs below assume no error in ACK/NACK, i.e. no channel noise. In Sec.~\ref{sec:sim}, we consider the impact of a realistic 5G channel on performance. 


\vspace*{-0.2cm}
\subsection{Analog Beam Alignment}
\label{subsec:analogba}

For analog BA, the BS can scan one beam at a time. We adapt Hwang's Generalized Binary Splitting (GBS), which is an interactive GT scheme that is asymptotically optimal in terms of the number of tests when an upper bound on the number of defectives is given \cite{hwang1972method}.

Our proposed analog GT-based BA (AGTBA) algorithm (Algorithm~\ref{alg:1}) has three inputs, $\Aset$ the angular intervals for data transmission (a.k.a data beam. codebook), $N = |\Aset|$ the number of beams in $\Aset$, and $M$ the number of paths. Data beam codebook $\Aset$ is obtained by dividing $[0, 2\pi)$ interval into equal angular intervals with beamwidth $\omega$.

Following the rationale of \cite{hwang1972method}, AGTBA performs an exhaustive search over $\Aset$ if the number of paths is more than half of the number of angular intervals in $\Aset$. If not, AGTBA forms a subset, $\Gset$, of $\Aset$ with $|\Gset|= 2^\alpha$ angular intervals, in which $\alpha$ is the size adjustment given in Algorithm~\ref{alg:1}. Then, AGTBA scans the beam whose angular region covers the union of the angular intervals in $\Gset$. If the scan returns NACK, it removes the set $\Gset$ from $\Aset$, updates $N$ and $M$, and performs AGTBA with these updated variables to select the rest of the scanning beams. However, if the scan returns an ACK, the algorithm selects the next set of scanning beams by performing Bisection-Search($\Gset$), which performs the bisection \cite{hussain2017throughput} on the set of angular intervals, $\Gset$, to find the one guaranteed path and returns the set of angular intervals, $\Gset'\subset \Gset$ which correspond to NACK responses.
Next, the algorithm removes $\Gset'$ from $\Aset$, updates $N$ and $M$, and performs AGTBA with the new set of variables to select the rest of the scanning beams until it finds $M$ AoAs or finishes scanning $N$ angular intervals. 
\vspace*{-0.35cm}
\begin{algorithm}
\DontPrintSemicolon
\SetNoFillComment

\caption{AGTBA$(\Aset, M, N)$ }\label{alg:1}
\lIf{$N \leq 2M -2$}
{
Exhaustive search}
\Else{
    \lIf{ M==1}
    {
    $\alpha = \floor{ \log_2 \frac{N}{2}}$
    }
    \lElse
    {$\alpha = \floor{ \log_2 \frac{N-M+1}{M} }$
    }
Form a beam by $2^\alpha$ angular intervals in $\Gset$ and scan it\;

    \eIf{NACK }
    {

        AGTBA($\Aset \setminus \Gset$, $M$, $N-2^\alpha$)\;
    }
    {
     $\Gset'$ =  Bisection-Search($\Gset$)\;
        AGTBA($\Aset \setminus \Gset'$, $M-1$, $N-|\Gset'|$);
    }
}
\end{algorithm}
\vspace*{-0.45cm}

Note that there is a modification for $M=1$ because the optimal BA procedure for $M=1$ is bisection search \cite{hussain2017throughput}; however, the GBS algorithm does not lead to bisection for $M=1$. This is because the GBS algorithm treats $M$ as an upper bound, so it confirms that there is a path by testing all of $\Aset$ once. However, $M$ is not an upper bound for BA since we need to detect at least one path between UE and BS. To fix this issue, we have modified GBS in Alg.~\ref{alg:1}, and have added the condition that $\alpha = \log_2N/2$ when  $M=1$.

\vspace*{-0.1cm}
\subsection{Hybrid Beam Alignment}
\label{subsec:hybridba}
Next, we consider hybrid BA, where the BS can check $N_{\rm RF}$ scanning beams at each time slot. Here, we provide BA procedures when $N_{\rm RF} = 2$ and comment on extensions to $N_{\rm RF} >2$ at the end of this section. The generalization of the proposed methods to more than two RF-chains and their performance analysis is left for future publication due to space constraints.

We provide modifications to AGTBA that allow for parallel testing, denoting these algorithms as GT-based BA (HGTBA). We consider the following three algorithms in the order of increasing complexity. The first one divides the optimization into two sub-problems and solves them in parallel.

\begin{figure*}[b]
\centering
\vspace*{-0.3cm}
\begin{subfigure}{0.31\linewidth}
    \centering
    \includegraphics[width=0.85\textwidth]{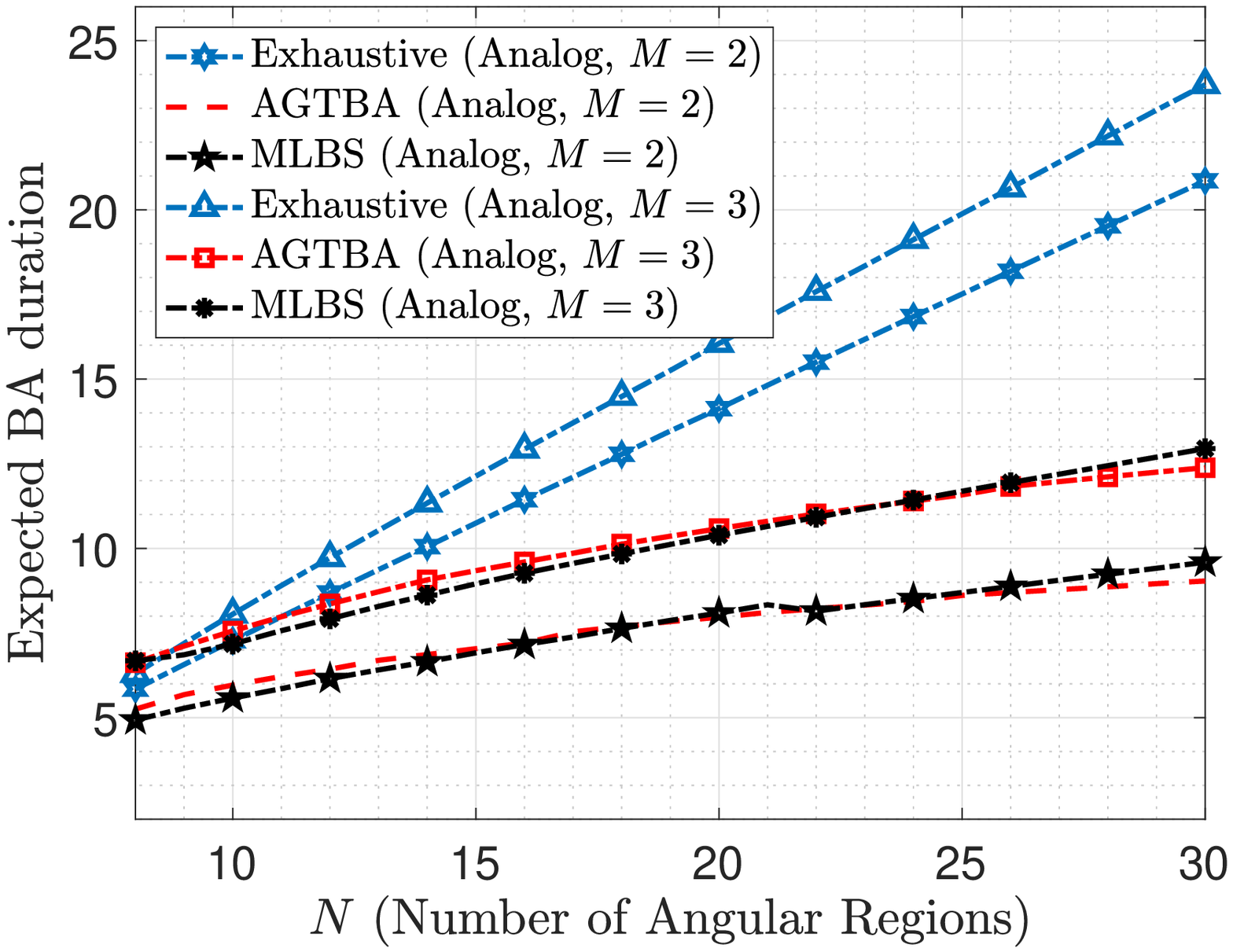}
    \caption{}
    \label{fig:gtmlbs}
\end{subfigure}
\begin{subfigure}{0.31\linewidth}
    \centering
    \includegraphics[width=0.85\textwidth]{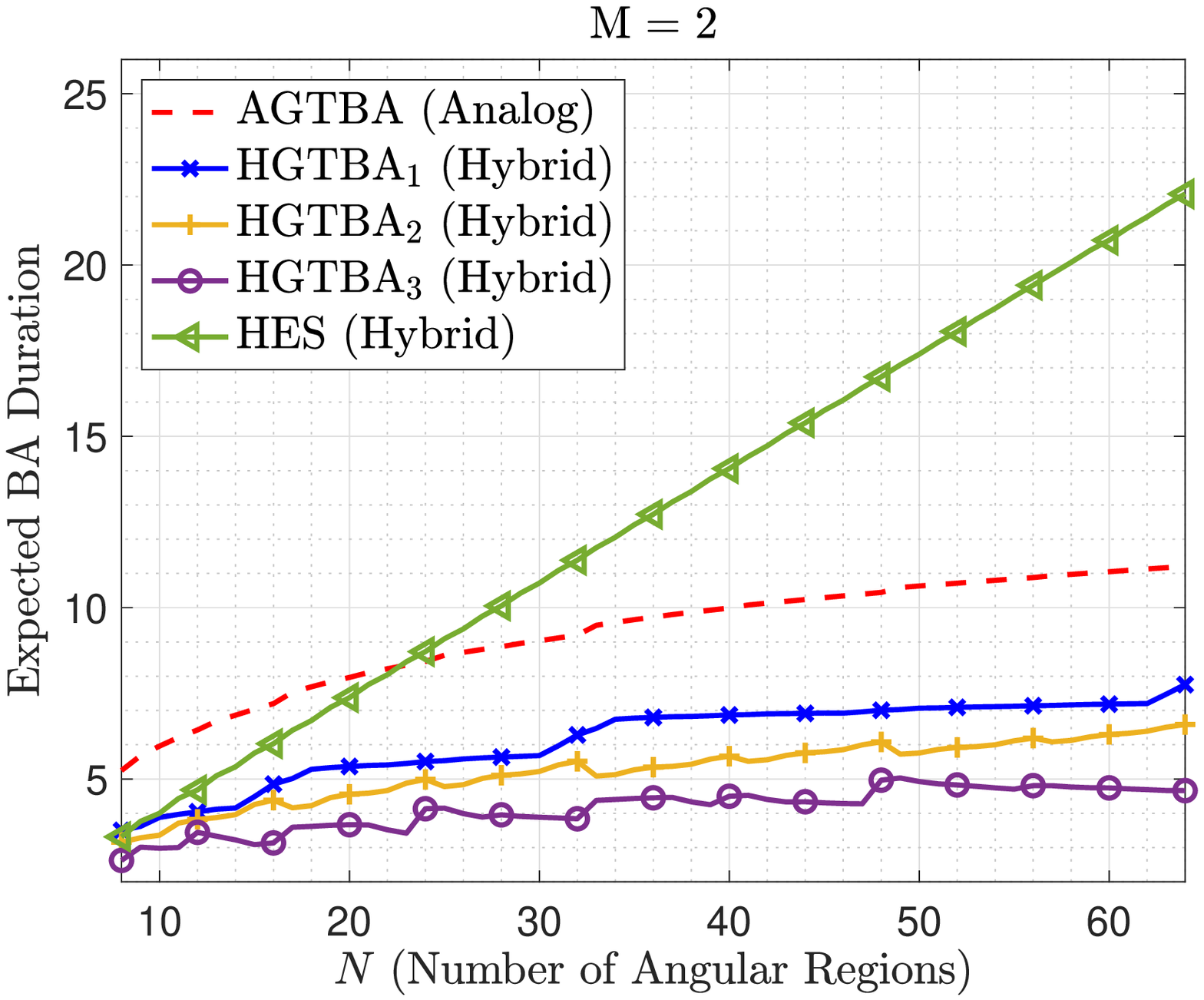}
    \caption{}
    \label{fig:p0m2}
\end{subfigure}
\begin{subfigure}{0.31\linewidth}
    \centering
    \includegraphics[width=0.85\textwidth]{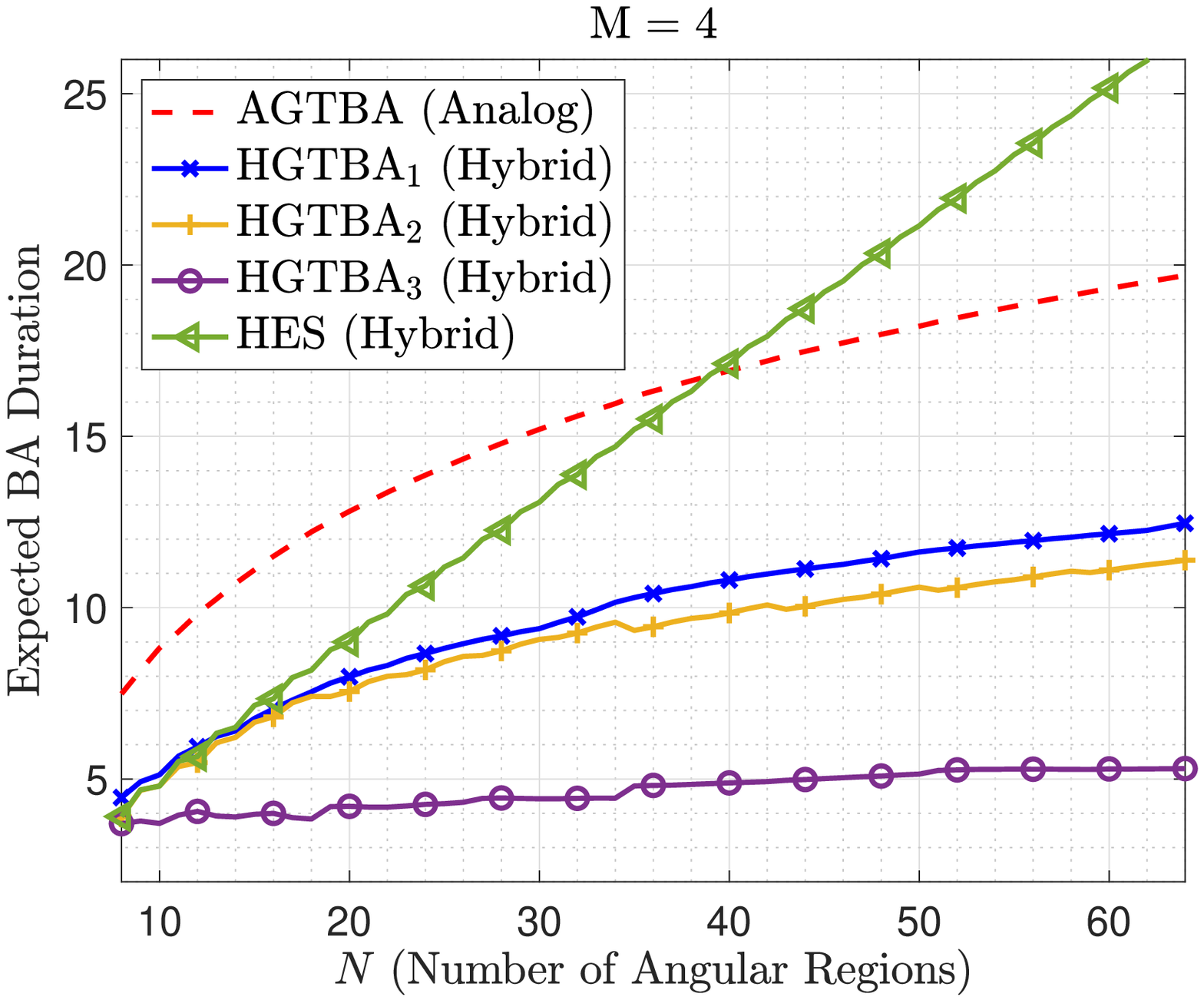}
    \caption{}
    \label{fig:p0m4}
\end{subfigure}
\vspace*{-0.2cm}
\caption{Expected BA duration versus number of angular regions $N$ for a) Analog BA: MLBS \cite{aykin2019multi}, exhaustive search \cite{ barati2016initial} and AGTBA for $M=2,3$, b) proposed BA algorithms and HES for $M=2$, and c) proposed BA algorithms and HES for $M=4$.}
\vspace*{-0.5cm}
\label{fig:set1}
\end{figure*} 

\begin{figure*}[t]
\centering
\vspace*{-0.2cm}
\begin{subfigure}{0.31\linewidth}
    \centering
     \includegraphics[width=0.85\textwidth]{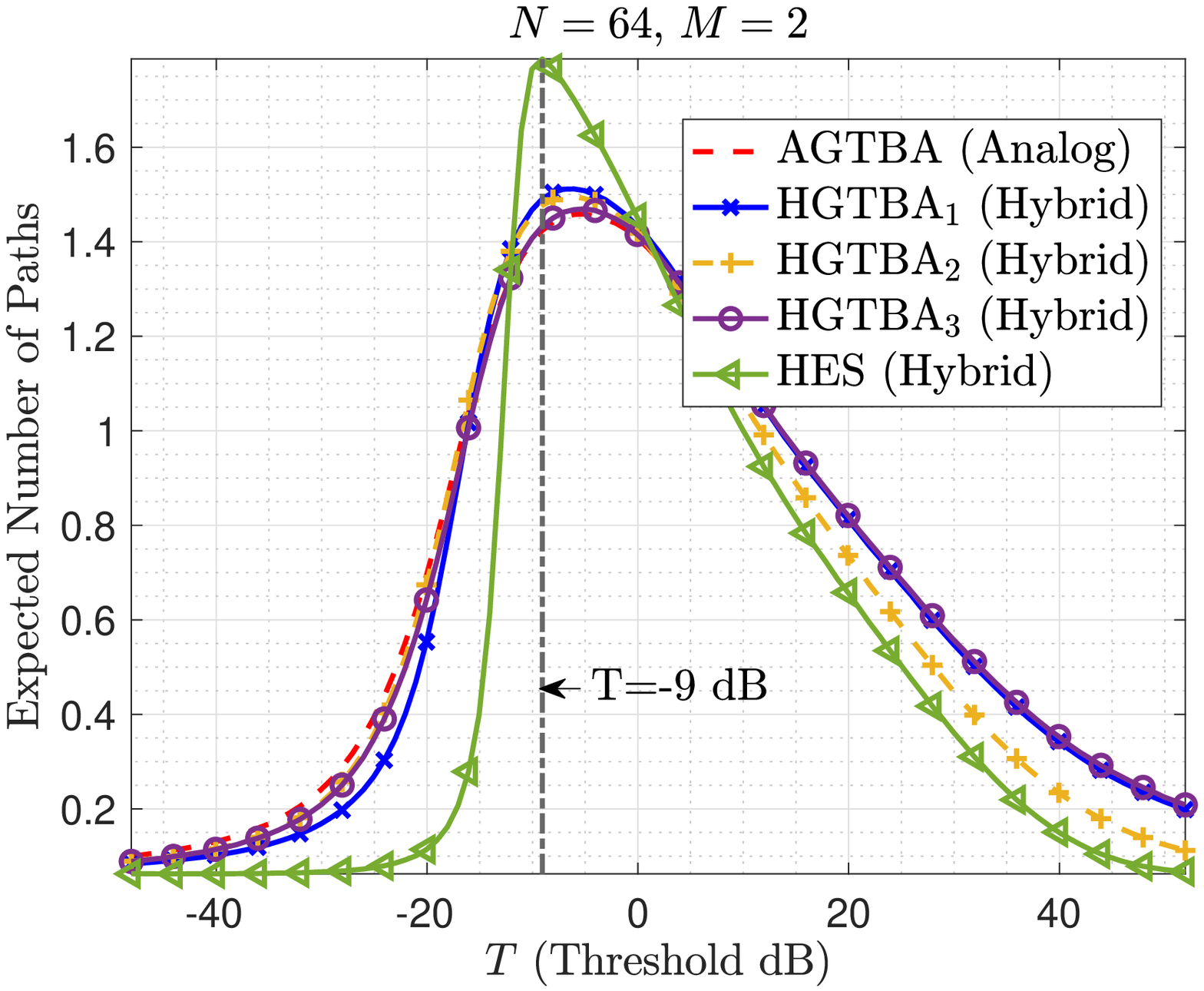}
    \caption{}
    \label{fig:mdm2pmd3.9e-01res}
\end{subfigure}
\begin{subfigure}{0.31\linewidth}
    \centering
     \includegraphics[width=0.85\textwidth]{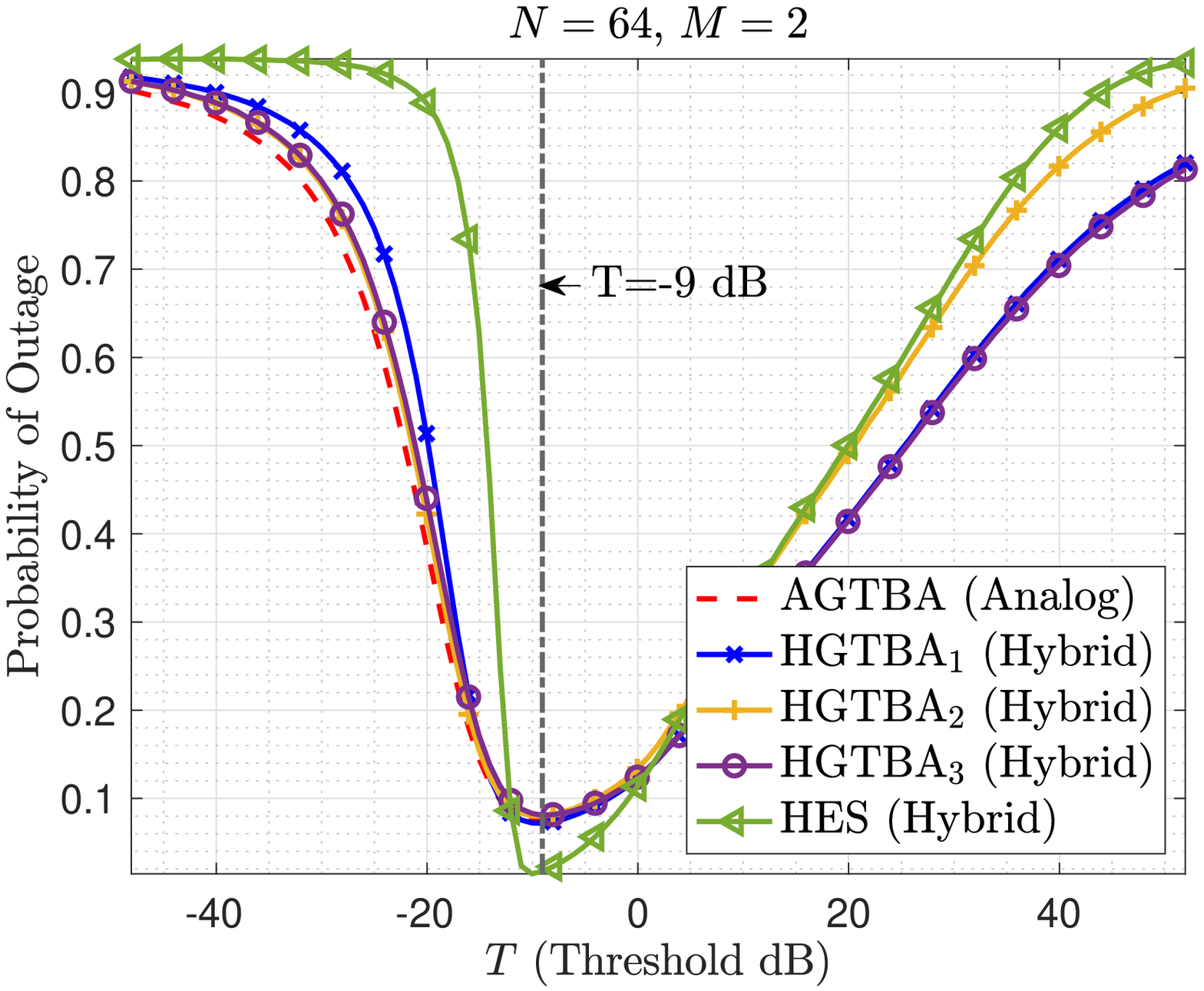}
    \caption{}
    \label{fig:blockagem2pmd3.9e-01res}
\end{subfigure}
\begin{subfigure}{0.31\linewidth}
    \centering
     \includegraphics[width=0.85\textwidth]{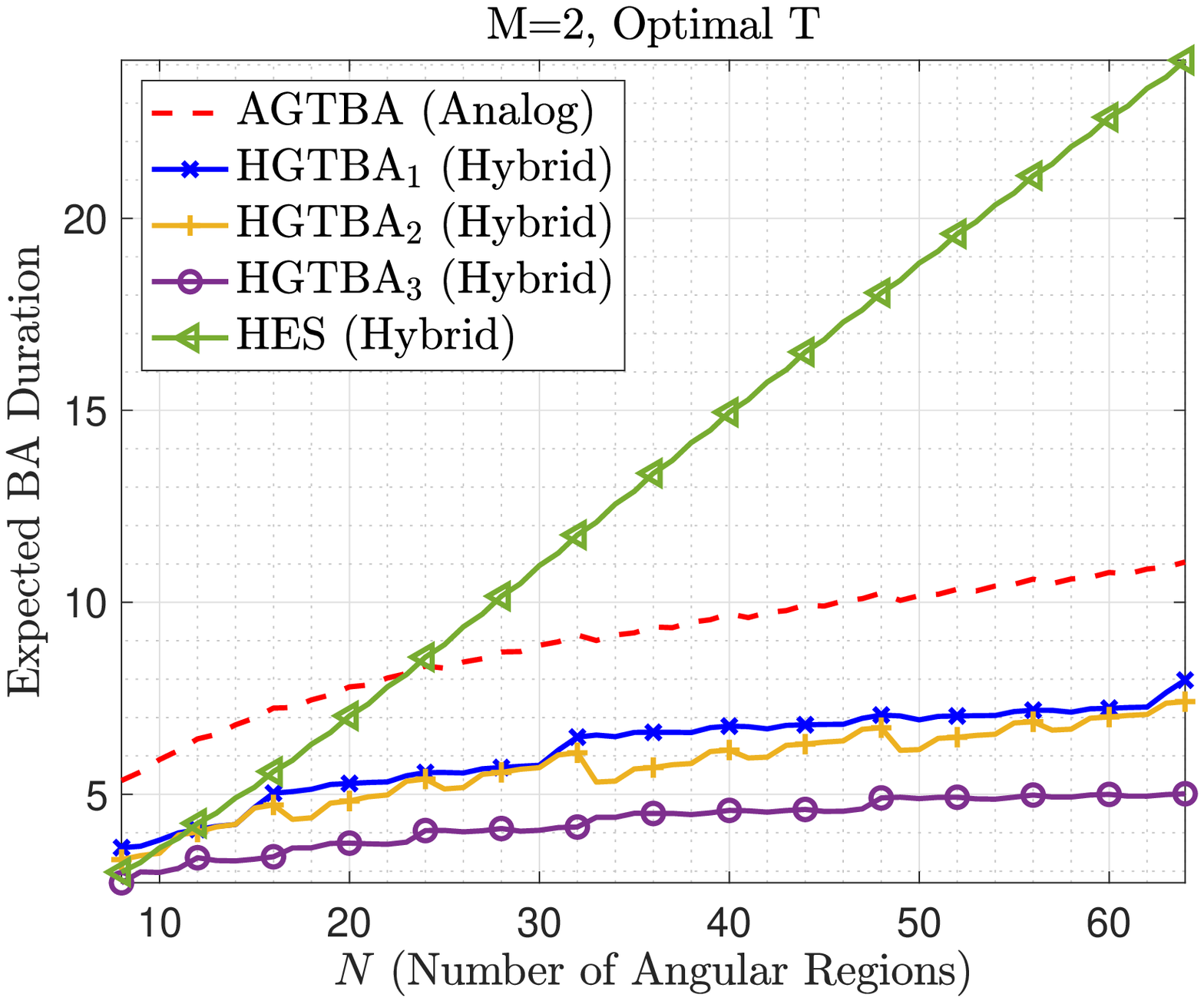}
    \caption{}
        \label{fig:2faaa}
\end{subfigure}

\vspace*{-0.2cm}
\caption{Comparison of the proposed algorithms and HES in terms of a) expected number of paths, and b) outage probability for varying thresholds, $M=2$, $N=64$, c) expected BA duration for the optimal threshold for different values of $N$, $M=2$.}
\vspace*{-0.5cm}
\label{fig:setthreshold}
\end{figure*}

\emph{Algorithm HGTBA$_1(\Aset,N,M)$}: We divide the data beam codebook $\Aset$ into two sets with roughly the same number of elements and perform AGTBA in parallel on the two sets considering $\approx{M/2}$ paths in each. After both AGTBA algorithms are completed, we remove the elements identified as  NACK during the searches, update $M$ and $N$, and repeat HGTBA$_1(\cdot)$ until $M$ AoAs are detected.

In the second hybrid algorithm, we jointly design the scanning beams of the parallel. searches discussed in HGTBA$_1$.  

\emph{Algorithm HGTBA$_2(\Aset,N,M)$}: Similar to AGTBA, when $N\leq 2M-2$, this algorithm performs an exhaustive search of the $N$ data beams with $N_{\rm RF} =2$ scanning beams at each time slot. Otherwise, it first forms two disjoint subsets of $\Aset$, namely $\Gset_1$ and $\Gset_2$ each consisting of $2^\alpha$ elements, where $\alpha$ is defined same as AGTBA (Note that $\Gset_2$ might have less than $2^\alpha$ elements depending on the size of $\Aset$). Here, $\Gset_1$ and $\Gset_2$ correspond to the scanning beams used for the first and second RF-chain, respectively. We can have one of the following: \emph{i)} Both scanning beams lead to ACKs. Then the algorithm selects the next scanning beams for the first (second) RF-chain by performing the bisection search over $\Gset_1$ ($\Gset_2$) to find one path. Once the search is over, the algorithm removes all the elements of $\Gset_1$ and $\Gset_2$ that are identified by NACKs from $\Aset$, updates $N$ and $M$, and performs HGTBA$_2(\cdot)$ with the new set of variables. \emph{ii)} One of the scanning beams or both lead to NACKs. Then the algorithm removes the corresponding sets $\Gset_i$s from $\Aset$, updates $N$, and performs HGTBA$_2(\cdot)$ with the new set of variables until it detects $M$ AoAs are detected or scans $N$ angular intervals.

We note that HGTBA$_2$ does not utilize every ACK information. When only one ACK is received, HGTBA$_2$ does not continue the search to find a path in the scanned interval. Hence, we propose our third algorithm in which every ACK results in bisection searching its corresponding angular interval.

\emph{Algorithm HGTBA$_3(\Aset,N,M)$}: This algorithm starts similar to HGTBA$_{2}(\cdot)$. However, for $N > 2M-2$, after forming $\Gset_1$ and $\Gset_2$ and performing the scans, if the response to only one of the scanning beams is ACK, the algorithm proceeds as follows. For the set resulting in ACK, the algorithm determines the scanning beams using a bisection search to find one path by using one RF-chain. Meanwhile, it uses other RF-chain to find a path by performing modified AGTBA (MAGTBA)
in which 
line 10 in Alg.~\ref{alg:1} is removed because AGTBA recursive call is not desired here. Also, since the set $\Aset \setminus \{\Gset_1\cup\Gset_2\}$ does not necessarily include a path, MAGTBA returns $p=1$ if there was a path; otherwise, $p=0$. 
Once the scans are over, the algorithm removes all the elements that are identified as NACK from $\Aset$, updates $N$ and $M$, and performs HGTBA$_3(\cdot)$ again. The pseudocode of this algorithm is provided in Algorithm~\ref{alg:2}.



For generalization to $N_{\rm RF} >2$ RF-Chains, one can create $N_{\rm RF}$ subgroups of size $\alpha'$, where $\alpha' = \alpha -k$ for the smallest positive k which satisfies $2^{\alpha'}N_{\rm RF} \leq N$ and proceed similarly as in all of the discussed algorithms. However, other designs are also possible and due to space constraint, the comparison between different methods for $N_{\rm RF}>2$ is left for future work.

\vspace*{-0.2cm}
\section{Simulations}
\label{sec:sim}
\vspace*{-0.1cm}


\subsection{Noiseless Case}
We first provide a simulation of the proposed analog and hybrid BA methods assuming no channel noise, i.e, no error in ACK/NACK. 
We average over $10^6$ realizations of the $M$ AoAs. Fig \ref{fig:gtmlbs} shows the comparison of AGTBA, and the state-of-the-art analog BA methods for $M$ paths, namely multi-lobe beam search (MLBS) algorithm \cite{aykin2019multi}, and exhaustive search for different values of $N$ and $M$. We have only considered $N\leq 30$ as the MLBS running time and memory requirement grow exponentially with N.  
For small values of $N$, AGTBA has performance similar to MLBS. However, as $N$ increases, AGTBA starts to outperform MLBS.
Moreover, AGTBA has lower computational complexity compared to MLBS which is due to the necessity of forming a decision tree. 

\vspace*{-0.35cm}
\begin{algorithm}
\DontPrintSemicolon
\SetNoFillComment

\caption{HGTBA$_3(\Aset, M, N)$}\label{alg:2}
\lIf{$N \leq 2M -2$}
{
  Exhaustive Search 
    }
\Else{
     \lIf{ M= 1}
    {$\alpha = \floor{ \log_2 \frac{N}{2}}$
    }\lElse{$\alpha = \floor{ \log_2 \frac{N-M+1}{M} }$}
    Form any disjoint sets $\Gset_1$ and $\Gset_2$ of size $2^\alpha$\;

    \uIf{ both NACK}
    {
        HGTBA$_3(\Aset \setminus \{\Gset_1 \cup \Gset_2\}, M, N-|\Gset_1|- |\Gset_2|)$\;
    }
    \uElseIf{both ACK}
    {
     $\Gset_i'$ =  Bisection-Search ($\Gset_i$) for $i$ in $\{1,2\}$ \;
    HGTBA$_3(\Aset \setminus \{\Gset_1' \cup \Gset_2' \},M-2, N-|\Gset_1'|- |\Gset_2'| )$

    }
        \Else
    {
     $\Cset_1, \Cset_2 =$ The set that resulted in ACK \& NACK\;
     $\Cset_1'$ =  Bisection-Search ($\Cset_1$)\;
     $\Gset_3'$, $p =$  MAGTBA$(\Aset \setminus \{\Gset_1 \cup \Gset_2\}, M-1, N-|\Gset_1|- |\Gset_2| )$\;
    HGTBA$_3(\Aset \setminus \{\Cset_1' \cup \Cset_2 \cup \Gset_3'\},M-1-p, N-|\Cset_1'|- |\Cset_2|-|\Gset_3'|)$
    }
     }
\vspace*{-0.2cm}

\end{algorithm}

\vspace*{-0.5cm}

We compare the performance of HGTBA algorithms in  Fig.~\ref{fig:p0m2} and  Fig.~\ref{fig:p0m4} along with hybrid exhaustive search (HES), in which the BS applies exhaustive search scanning $N_{\rm RF} = 2$ angular intervals at a time. To compare with analog BA, we plotted the performance of AGTBA. As expected from Sec.~\ref{subsec:hybridba}, we observe that the HGTBA$_3$ algorithm has the least expected BA duration. Moreover, in  Fig.~\ref{fig:p0m2}, HGTBA$_3$ reduces $E[T_{\rm BA}]$ by a factor of two compared to AGTBA when $M=2$. Based on Fig.~\ref{fig:p0m4}, the reduction in $E[T_{\rm BA}]$ is further increased to a factor of three when $M=4$. We also observe that when $N$ is small, HGTBA$_1$ and HGTBA$_2$ algorithms perform similarly to HES. However, for large values of $N$, all our proposed algorithms outperform HES.

\vspace*{-0.2cm}

\subsection{5G Network Simulations}
\vspace*{-0.15cm}

\emph{Network Model:} 
We investigate the performance of the proposed algorithms under a realistic 5G mmWave network model. We consider a single-cell uplink scenario with a BS at the origin with height $10$ m and a UE with height $2$ m operating at 28 GHz with $57.6$ MHz bandwidth. The UE is placed uniformly in a ring around the BS with inner and outer radii of $10$ to $200$ meters. We assume that the UE has one antenna while the BS is using a uniform linear array with $64$ antennas. The UE transmits a BA symbol at each time slot with $20$ dBm power. We consider the clustered channel model \cite{akdeniz2014millimeter} with an additive Gaussian noise consisting of $M$ clusters each  with one path, where the channel gain vector, $\mathbf{h}\in \mathbb{C}^{N_{\rm Rx}}$ can be written as 
\begin{align}
\label{eq:cch}
  \hbf  = \sum_{m=1}^{M} \sqrt{N_{\rm RX}}h_m\dbf_r(\psi_m) 
\end{align}
where $\psi_{m}$ is AoA of path $m$ generated uniformly in $(0,2\pi]$, $h_m$ is the $m^{\rm th}$ path's channel gain including its path loss and small scale fading generated based on 3GPP channel parameters \cite{3gppchannel}, and $\dbf_r(\cdot)$ is the array response vector of the BS. We assume that the channel during BA is static. 

Similar to \cite{hunter2008transmission}, we use \textit{sectored antenna} model from \cite{ramanathan2001performance}. We assume given $N_a$ \textit{active} number of antennas for an RF-chain, we can form a contiguous beam $\Phi$ of width $2\pi/N_a$ with power gain of $N_a$ for that RF-chain. Thus, if a path is included in $\Phi$, the beamforming gain is $N_a$; otherwise, the gain is zero. Note that in our proposed algorithms, $N_a$ needs to vary at each time slot for each RF-chain since the width of the scanning beams changes.




Based on the beamformed signal, we need to decide whether the BS's measurement results in ACK or NACK. We use an energy detection approach, in which the energy of the measured signal, $P_{\rm RX}$, is normalized by the beamforming gain $N_a$ 
and compared with a threshold $T$. 
If $P_{\rm RX}/N_a\geq T$ we consider it ACK, else NACK. The normalization is to ensure the threshold value independent of the beamwidth. The implementation of the algorithms and the channel simulation can be found on Github \footnote{https://github.com/OzleemYildiz/Beamforming}.

\emph{Simulation results:}
We first investigate the impact of different thresholds on the performance of the proposed algorithms. To this end, we plot the average performance over $10^6$ independent channel realizations for different values of $T$ when $N=64$ and $M=2$. In Fig.~\ref{fig:mdm2pmd3.9e-01res}, we show the expected number of detected paths in a 5G network. We observe that for low and high threshold values, the expected number of detected paths is close to zero and the optimal performance happens in between. For low threshold values, the false alarm probability (probability of detecting a path while no path is included in the beam) is high so measurements lead to ACK with high probability. On the other hand, for high threshold values, the missdetection probability (probability of not detecting a path while at least a path is included in the beam) is high so measurements are likely to be NACK. Fig~\ref{fig:mdm2pmd3.9e-01res} and Fig~\ref{fig:blockagem2pmd3.9e-01res} suggest that the slope of the transition in the number of paths detected in the false alarm probability region is sharper than in the missdetection probability region. Thus, we conclude that the proposed BA methods are more sensitive to a false alarm than missdetection probability. 
Similar observations can be made from Fig.~\ref{fig:blockagem2pmd3.9e-01res}, where we have plotted the outage (missing all paths) probability of the BA methods. We observe that although HES has marginally better performance for the optimal threshold in terms of the expected number of detected paths and blockage 
rate, it is less robust (sharper transition) to variations of the threshold compared to the proposed BA methods. Finally, as illustrated in Fig.~\ref{fig:2faaa}, for optimized thresholds, the proposed BA methods lead to a lower expected BA duration than HES and the performance in terms of expected BA duration are similar to the noiseless case in Fig~\ref{fig:set1}.

\section{Conclusion}

In this paper, we have investigated interactive hybrid BA in the uplink, where the channel between UE and BS consists of multiple paths. We have proposed a general GT-based framework through which we have provided novel GT-based analog and hybrid BA strategies. We have shown through simulations that the proposed BA strategies lead to reduced expected BA duration compared to the state-of-the-art methods both in idealized noiseless and realistic 5G settings.

\bibliographystyle{IEEEtran}
\bibliography{ref}

\end{document}

%% file: others/com.tex
\newcounter{newenumi}
\usepackage[normalem]{ulem} 

\newcommand{\citep}[1]{\cite{#1}}

\def\beq{\begin{equation}}
\def\eeq{\end{equation}}
\def\beqa{\begin{eqnarray}}
\def\eeqa{\end{eqnarray}}
\def\beqan{\begin{eqnarray*}}
\def\eeqan{\end{eqnarray*}}

\DeclareMathOperator*{\argmin}{arg\,min}

\theoremstyle{definition}

\def\Cset{{\cal C}}

\def\Aset{{\cal A}}

\newcommand{\dbf}{\mathbf{d}}

\newcommand{\hbf}{\mathbf{h}}

\def\Gset{{\cal G}}